\documentclass[twocolumn, showpacs,preprintnumbers,amsmath,amssymb,superscriptaddress, longbibliography]{revtex4}

\usepackage{graphicx}
\usepackage{dcolumn}
\usepackage{bm}
\usepackage{xfrac}
\usepackage[detect-all]{siunitx}
\newcommand{\ex}{E$^{*}$} 
 
\newcommand{\al}{$\alpha$}
\newcommand{\als}{$\alpha$ } 
\newcommand{\ca}{$^{40}$Ca}
\newcommand{\cas}{$^{40}$Ca } 

\newcommand{\carbs}{$^{12}$C } 

\newcommand{\oxs}{$^{16}$O } 

\newcommand{\neons}{$^{20}$Ne } 

\newcommand{\mgs}{$^{24}$Mg }
\newcommand{\li}{$^{6}$Li}
\newcommand{\lis}{$^{6}$Li }
\newcommand{\fos}{$^{30}$P}
\newcommand{\glin}{$^{26}$Al}
\newcommand{\na}{$^{22}$Na}
\newcommand{\proc}{$\%$}
\newcommand{\procs}{$\%$ }
\newcommand{\alm}{A$_{L}$}
\newcommand{\cyclotron}{Cyclotron Institute, Texas A$\&$M University, College Station, Texas
77843}

\begin{document}

\preprint{APS/123-QED}

\title{Alpha Conjugate Neck Structures in the Collisions\\ of  35 MeV/nucleon
$^{40}$Ca with $^{40}$Ca}
\author{K.~Schmidt}
\email{katarzyna.schmidt@us.edu.pl} 
\affiliation{\cyclotron}
\affiliation{Institute of Physics, University of Silesia, Katowice, Poland}

\author{X.~Cao} 
\affiliation{Shanghai Institute of Applied
Physics, Chinese Academy of Sciences, Shanghai 201800, China}
\affiliation{\cyclotron}

\author{E.~J.~Kim} 
\affiliation{\cyclotron}
\affiliation{Division of Science Education,
Chonbuk National University, Jeonju 561-756, Korea}

\author{K.~Hagel}
\affiliation{\cyclotron}

\author{M.~Barbui} 
\affiliation{\cyclotron}

\author{J.~Gauthier} 
\affiliation{\cyclotron}

\author{S.~Wuenschel} 
\affiliation{\cyclotron}

\author{G.~Giuliani}
\affiliation{\cyclotron}
\affiliation{Laboratori Nazionali del Sud,
INFN,via Santa Sofia, 62, 95123 Catania, Italy}

\author{M.~R.~D.~Rodrigues} 
\affiliation{\cyclotron}
\affiliation{Instituto de F\'isica,
Universidade de S\~ao Paulo,Caixa Postal 66318, CEP 05389-970, S\~ao Paulo, SP,
Brazil} 

\author{H.~Zheng}
\affiliation{\cyclotron}
\affiliation{Laboratori Nazionali del Sud,
INFN,via Santa Sofia, 62, 95123 Catania, Italy}

\author{M.~Huang}%
\affiliation{\cyclotron}
\affiliation{College of Physics and Electronics
information, Inner Mongolia University for Nationalities, Tongliao, 028000,
China}

\author{N.~Blando}%
\affiliation{\cyclotron}

\author{A.~Bonasera} 
\affiliation{\cyclotron}
\affiliation{Laboratori Nazionali del Sud,
INFN,via Santa Sofia, 62, 95123 Catania, Italy}

\author{R.~Wada}
\affiliation{\cyclotron}

\author{C.~Botosso}
\affiliation{\cyclotron}

\author{G.~Liu}
\affiliation{Shanghai Institute of Applied Physics, Chinese Academy of Sciences,
Shanghai 201800, China}

\author{G.~Viesti}
\affiliation{Dipartamento di Fisica dell'Universit$\grave{a}$ di Padova and INFN Sezione di
Padova, Italy}

\author{S.~Moretto}
\affiliation{Dipartamento di Fisica dell'Universit$\grave{a}$ di Padova and INFN Sezione di
Padova, Italy}

\author{G.~Prete}
\affiliation{INFN Laboratori Nazionali di Legnaro, Italy}

\author{S.~Pesente}
\affiliation{Dipartamento di Fisica dell'Universit$\grave{a}$ di Padova and INFN Sezione di
Padova, Italy}

\author{D.~Fabris}
\affiliation{Dipartamento di Fisica dell'Universit$\grave{a}$ di Padova and INFN Sezione di
Padova, Italy}

\author{Y.~El~Masri}
\affiliation{Universit´e Catholique de Louvain, Louvain-la-Neuve, Belgium}

\author{T.~Keutgen}
\affiliation{Universit´e Catholique de Louvain, Louvain-la-Neuve, Belgium}

\author{S.~Kowalski}
\affiliation{Institute of Physics, University of Silesia, Katowice, Poland}

\author{A.~Kumar}
\affiliation{Nuclear Physics Laboratory, Department of Physics, Banaras Hindu
University, Varanasi, India}

\author{G.~Zhang} 
\affiliation{\cyclotron}
\affiliation{Shanghai Institute of Applied
Physics, Chinese Academy of Sciences, Shanghai 201800, China}

\author{J.~B.~Natowitz}
\affiliation{\cyclotron}

\date{\today}

\begin{abstract} 
The de-excitation of alpha-conjugate nuclei produced in reactions of 35 MeV/nucleon \cas with \cas has been investigated. 
Particular emphasis is placed on examining the dynamics of collisions leading to projectile-like fragment exit channels. 
A general exploration of the reaction systematics reveals the binary dissipative character of the collisions and a hierarchy effect 
similar to that seen for heavier systems. Investigation of the subset of events characterized by a total \al-conjugate mass (\als particles plus 
\al-conjugate fragments) equal to 40 and atomic number equal to 20 reveal a dominance of \al-conjugate exit channels. 
The hierarchy effect for these channels leads to the production of \al-clustered neck structures with potentially exotic geometries and properties.
\end{abstract}

\maketitle

\section{\label{sec:first}Introduction} 
Nuclei are normally treated as
consisting of fermions.  However, in medium correlations and the strong binding
of the \als particle can lead to situations in which an \als cluster picture
can be employed to understand nuclear structure and decay
properties~\cite{beck2010clusters, ikeda1968systematic, freer2007clustered,
von2006nuclear}. Both theoretical calculations and experimental observations
provide strong support for the \als clustered nature of light \al~-~conjugate
(even-even N=Z) nuclei~\cite{kanada2001structure, epelbaum2012structure,
johnson2009extreme}.
Loosely bound states with excitation energies near the alpha emission thresholds
states may be a manifestation of the tendency of low density low temperature nuclear matter to undergo Bose  condensation 
~\cite{fukui2016probing,
funaki2008alpha, tohsaki2001alpha, ropke1998four, umar2010microscopic}.  
For example, the 7.65 MeV Hoyle-state in \carbs, important for the  solar 3\als capture process~\cite{hamada2013observation} is known to possess a large 
radius~\cite{zinner2013comparing}, which could allow  the \als particles to retain their quasi-free characteristics.\\
\indent
The role of \als clusters in reaction dynamics is itself an interesting topic. Cluster effects are often seen in transfer reactions involving light 
nuclei ~\cite{hodgson2003cluster}. Studies of more violent collisions of \als conjugate nuclei might reveal important effects of these correlations 
on the collision dynamics and in determination of the reaction exit channels. Given that near Fermi energy nuclear collisions can drastically modify 
the temperatures, densities and cluster properties of nucleonic matter, the possibility that short-lived Bose Condensates might be fleetingly produced 
in such collisions is an intriguing idea. Recently the emission of three \als from the Hoyle state has been characterized for \carbs produced in several 
different reactions ~\cite{raduta2011evidence, manfredi2012alpha, kirsebom2012improved, messina2015}. Results for the ratio of simultaneous to sequential 
de-excitation differ and the influence of medium or proximity effects on the de-excitation modes of that state in complex reactions remains an open question. 
The authors of  reference ~\cite{borderie2016probing} have argued that enhanced \als emission occurs during the thermal expansion of \oxs, \neons and 
\mgs projectile-like fragments produced in 25 MeV/nucleon, \cas ~+~\carbs collisions, reflecting the \al-conjugate nature of the parent fragments. 
Signatures of and possible evidence for Bose Einstein condensation and Fermi quenching in the decay of hot nuclei produced in 35MeV/nucleon \cas~+~\cas 
collisions have been discussed in references ~\cite{marini2016signals, zheng2012higher, zheng2013coulomb}. Evidence of cluster effects in the dynamics at 
much higher energies were reported in reference ~\cite{mathews1982inclusion}.\\  
\indent 
To pursue the question of the effects of \al-like correlations and clustering in collisions between \al-conjugate nuclei we have embarked on a program of 
experimental studies of such collisions at and below the Fermi energy using the NIMROD-ISiS array at TAMU ~\cite{wuenschel2009nimrod}. A dominating 
\als clustered nature of the colliding matter could manifest itself in the kinematic properties and yields of the \als conjugate products. While the 
granularity of our detection system is not sufficient for high resolution fragment and particle correlations, we are able to explore certain features of the 
reactions which lead to large cross sections for \als conjugate reaction products.\\
\indent In this paper we report results for a study of \als clusterization effects 
in mid-peripheral collisions of \cas~+~\cas at 35 MeV/nucleon. We first present some global observations and then focus on collisions in which excited 
projectile-like fragments disassemble into \al-conjugate products.  
\section{\label{sec:exp}Experimental Details} 
The
experiment was performed at Texas A$\&$M University Cyclotron Institute. \cas
beams produced by the $K500$ superconducting cyclotron  impinged on \cas targets
at the energy of 35~MeV/nucleon. The reaction products were measured using a
$4\pi$ array, NIMROD-ISiS (Neutron Ion Multidetector for Reaction Oriented
Dynamics with the Indiana Silicon Sphere)~\cite{wuenschel2009nimrod}  which
consisted of $14$ concentric rings covering from $3.6^{\circ}$ to $167^{\circ}$
in the laboratory frame. In the forward rings with $\theta_{lab}\leq
45^{\circ}$, two special modules were set having two Si detectors ($150$ and
$500$ $\mu m$) in front of a CsI(Tl) detector ($3-10\ cm$), referred to as
super-telescopes. The other modules (called telescopes) in the forward and
backward rings had one Si detector (one of $150$, $300$ or $500$ $\mu m$)
followed by a CsI(Tl) detector. The pulse shape discrimination method was
employed to identify the light charged particles with $Z\leq3$ in the CsI(Tl)
detectors. Intermediate mass fragments (IMFs), were identified with the
telescopes and super-telescopes using the ``$\Delta E-E$'' method. In the
forward rings an isotopic resolution up to $Z=12$ and an elemental
identification up to $Z=20$ were achieved. In the backward rings only
Z~=~\numrange[range-phrase = --]{1}{2}
particles were identified, because of the detector energy thresholds. In
addition, the Neutron Ball surrounding the NIMROD-ISiS charged particle array
provided information on average neutron multiplicities for different selected
event classes.  Further details on the detection system, energy calibration,
and neutron ball efficiency can be found in~\cite{wuenschel2009nimrod,
hagel2000light, wang2005tracing}. 

It is important to note that, for symmetric collisions in this energy range, the
increasing thresholds with increasing laboratory angle lead to a condition in
which the efficiencies strongly favor detection of projectile-like fragments
from mid-peripheral events. The modeling of these collisions using an
Antisymmetrized Molecular Dynamics (AMD) code~\cite{ono1999antisymmetrized,
ono2016cluster} coupled with the statistical code
GEMINI~\cite{charity1988systematics} as an afterburner, and applying the
experimental filter  demonstrates that this is primarily an effect of energy
thresholds.  
\section{\label{sec:general}General Characterization of the
Reactions} 
\begin{figure}
\includegraphics[width=\linewidth]{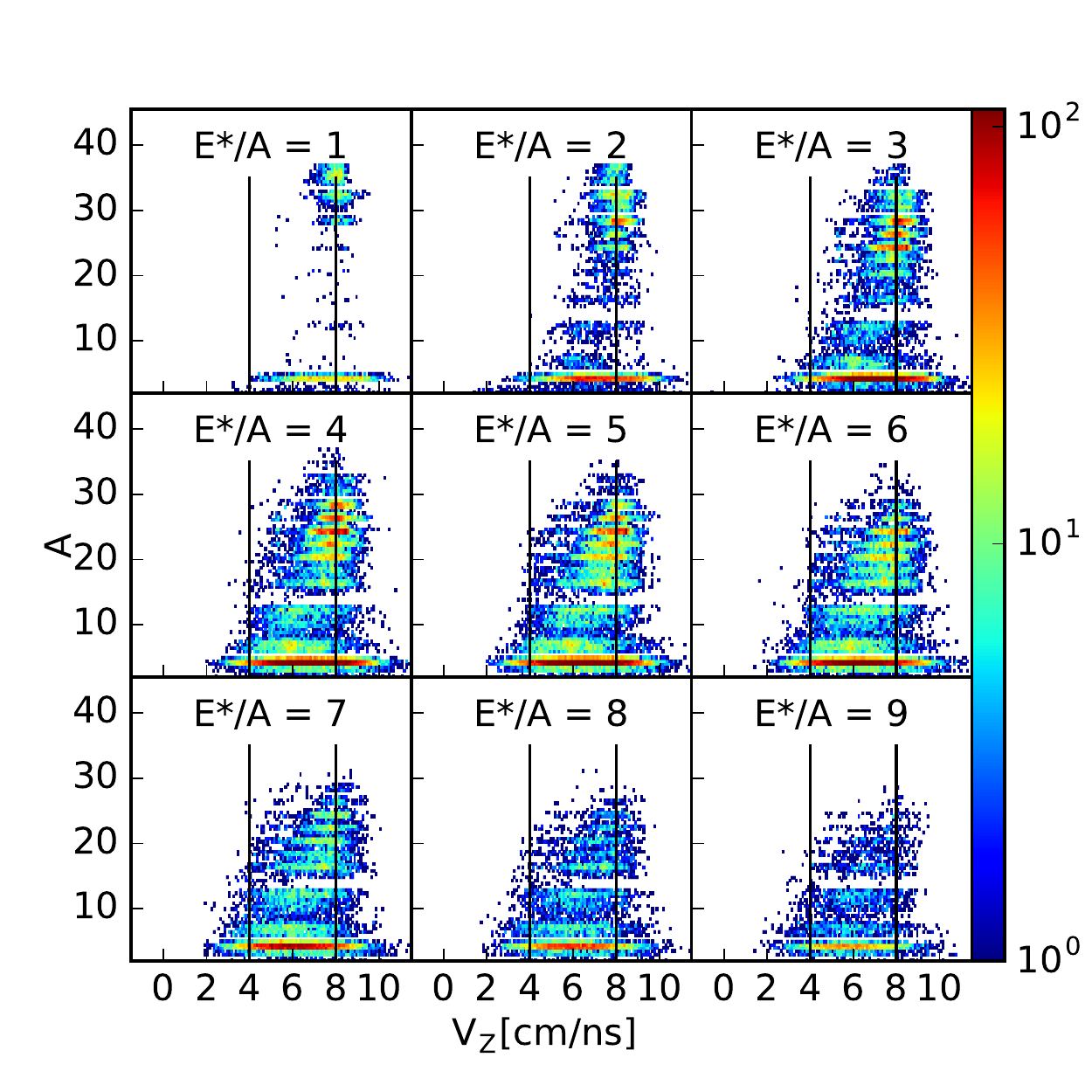}
\caption{\label{fig:fig1}}(Colour online.) Yields of the three heaviest
fragments in the event as a
function of the fragment parallel velocity in different windows of initial
apparent excitation energy \ex/A. The projectile velocity is 8.0[cm/ns]. The
c.m. velocity is 4.0[cm/ns].These two velocities are indicated by vertical lines
in each panel.  
\end{figure}

\begin{figure}
\includegraphics[width=\linewidth]{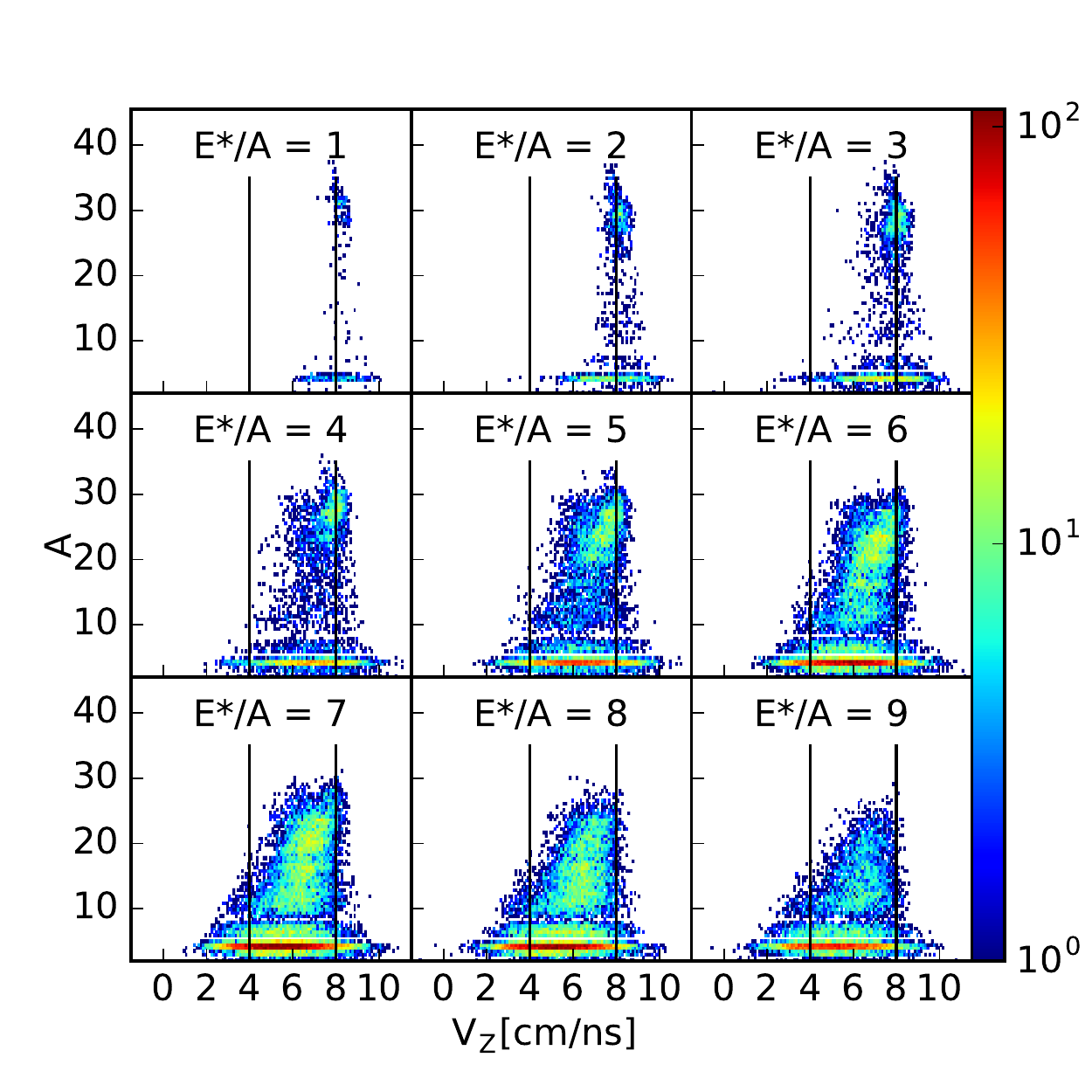}
\caption{\label{fig:fig2}}(Colour online.) Filtered AMD-GEMINI results, similar
as Fig.~\ref{fig:fig1}.
\end{figure}

Our previous study of the \cas + \cas at 35~MeV/nucleon  focused on
the multi-fragment exit channels and led to the conclusion that even the most
violent and most central collisions were binary in
nature~\cite{hagel1994violent}. Similar conclusions on the dominant binary
nature of reactions with 35~MeV/nucleon \mgs projectiles were reported by
Larochelle {\it{et al.}}~\cite{larochelle1996dependence}.\\ 
\indent We initiated the present analysis of the new data by reconstructing the 
``initial apparent excitation
energy'', \ex,  of the projectile-like fragments through calorimetry. \ex was
defined as the sum, for accepted particles, of the particle kinetic energies in
the frame of the total projectile-like nucleus (determined by reconstruction of
the mass and velocity of the primary excited nucleus from its de-excitation
products), minus the reaction Q-value. See equation~(\ref{eq:ex}).
\begin{eqnarray} 
E^{*} = \sum^{M}_{i=1}{K_{cp}(i) + M_{n}\langle K_{n}\rangle -
Q}. 
\label{eq:ex} 
\end{eqnarray} 
Here M is the total charged particle
multiplicity, K$_{cp}$(i) is the source frame kinetic energy of charged
particle~i, M$_n$ is the average neutron multiplicity, $\langle K_{n}\rangle$ is the
average neutron kinetic energy and Q is the disassembly Q value. For this
purpose the average kinetic energy of the neutrons was taken to be equal to the
average proton kinetic energy with a correction for the Coulomb barrier energy.
Average neutron multiplicities were determined by applying efficiency
corrections to the average neutron multiplicities observed with the neutron
ball~\cite{wang2005tracing}. For a compound nucleus this initial apparent
excitation energy would correspond to the energy available for statistical decay
of the primary nucleus.  We caution that given the binary nature of the
collisions studied, the de-exciting projectile-like nucleus is not necessarily a
fully equilibrated nucleus.  Nevertheless, this measure of energy deposition
into the systems studied can serve as a useful sorting parameter. For the
initial event selection we included all particles and fragments detected in an
event. As will be seen, this event selection is revised in subsequent sections
where we employ a more restrictive filtering to derive excitation energies.\\ 
\indent In Fig.~\ref{fig:fig1} the mass numbers, A, of the three heaviest fragments in each
event are plotted against their laboratory-frame parallel velocities for 1 MeV
increments in \ex/A. The favored detection of projectile-like species for all
windows is clearly seen in this figure. Most of the fragments have velocities
above the center of mass velocity, 4.0[cm/ns].  
Increasing excitation energy corresponds, at least
qualitatively, to decreasing impact parameter and increased collision violence.
This is manifested in the figure by the decrease in yields of the heaviest mass
products and increasing yields of lighter mass products as excitation increases.
At low excitation energies the majority of the heavier products have parallel
velocities near the beam velocity of 8.0[cm/ns]. The similar mean lab velocities
suggest that the lighter fragments are produced in by statistical de-excitation
of the initial projectile-like fragment. As the excitation energy increases, a
clear correlation between parallel velocity and fragment mass is observed. For these
excitations, corresponding to the region of mid-peripheral collisions, the
parallel velocity decreases as the fragment mass decreases.
This trend could reflect a greater degree of energy dissipation with decreasing impact parameter and/or the onset of neck emission 
~\cite{colin2003dynamical, baran2004neck, larochelle1997formation}. We shall return to this question. \\
Fig.~\ref{fig:fig2} shows the results of AMD-GEMINI calculations for this 
35~MeV/nucleon \cas + \cas system filtered using our experimental geometries and
thresholds. The AMD calculation~\cite{ono1999antisymmetrized, ono2016cluster}
followed the reaction until 300 fm/c after the collision.  The code
GEMINI~\cite{charity1988systematics} was employed as an afterburner to de-excite
the primary fragments. We note that the plots in Fig.~\ref{fig:fig2} look qualitatively
similar to those in Fig.~\ref{fig:fig1}. However at the lower excitation energies the AMD exhibit narrower velocity distributions and different 
yield distributions. This may be a manifestation of more transparency in the AMD collision than in the experiment ~\cite{wada2004reaction}.   
\section{\label{sec:plf}Selection of A~=~40, Z~=~20  PLF}
Our previous analyses of near Fermi energy collisions~\cite{hagel2000light,
wang2005tracing} indicate that significant proton emission occurs in the
earliest stages of the collision as the nucleon momentum distributions are
thermalizing, not in the later stage disassembly. To better characterize the
source of the light particles in the selected events we explored the Z~=~1 and
Z~=~2 light particle emission by carrying out both 2-source and 3-source
fits assuming that the observed light charged particle emission can be
attributed to primary sources moving in the laboratory frame, a projectile-like
source (PLF), a target-like source (TLF) and a  (virtual)  intermediate velocity
source (IV)  moving at a velocity~$\thicksim\sfrac{1}{2}$ the projectile
velocity~\cite{fuchs1994heavy}.  This latter source 
reflects nucleon-nucleon collisions occurring early in the process. In each
source frame the emission was assumed to have a Maxwellian distribution and each
of the sources is described by a source velocity, temperature, Coulomb barrier
and particle multiplicity~\cite{hagel2000light}. A comparison of the total
yields with those obtained from the source fits to the proton energy spectra
indicates  the proton  emission is low, with average multiplicities~$\thicksim$2
and is dominated by emission from an intermediate velocity source having an
apparent velocity of~$\thicksim\sfrac{1}{2}$ that of the projectile rather than
from later statistical  de-excitation. For this light symmetric system we expect
the same to be true for the neutrons. While the neutron kinetic energies are not
accessible in this experiment, the efficiency corrected neutron multiplicities
obtained using the neutron ball are similar to the proton multiplicities. \\ For
d and t emission the average multiplicities are much lower and about half the
particles are emitted from the IV source. The $^3$He emission was too low to
allow reasonable fits. To pursue our analysis we focus on events for which
A~=~40 and Z~=~20. However in this selection we have neglected both protons and
neutrons.
\section{\label{sec:stat}Tests of Statistical Behavior} 
\begin{figure} 
\includegraphics[width=\linewidth]{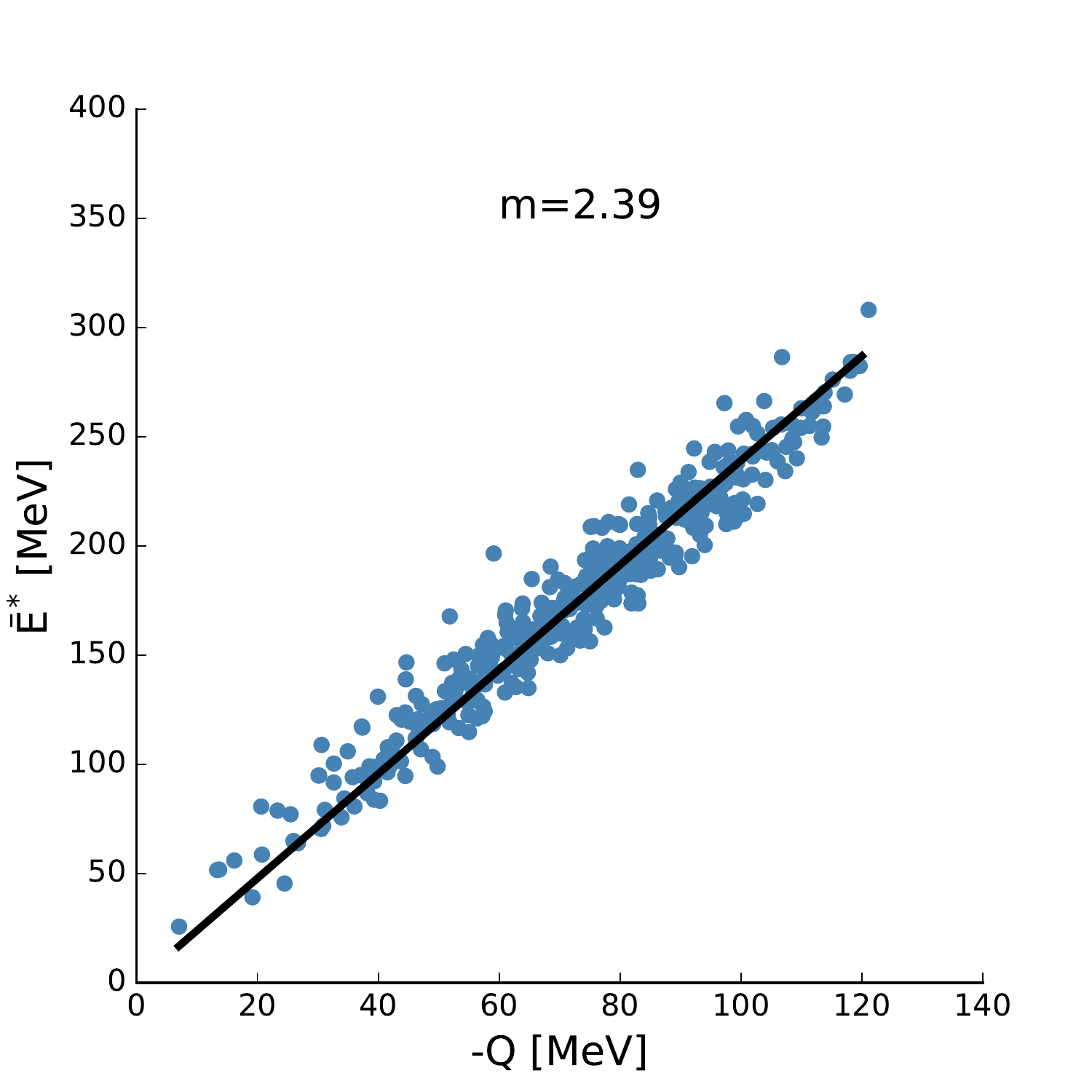}
\caption{\label{fig:fig3}}(Colour online.) Average excitation energy vs exit channel separation
energy for the de-excitation channels of A~=~40, Z~=~20 nuclei selected as
described in the text. Data are represented by small filled dots. The linear
least squares fit to the data is represented by the solid line.
\end{figure}

Horn and co-workers
suggested that the ratio of average excitation energy to the average exit
channel separation energy could be used as a test for statistical emission from
highly excited lighter nuclei
~\cite{larochelle1996dependence,
colin2003dynamical, 
baran2004neck, 
larochelle1997formation, 
fuchs1994heavy,
horn1996signatures}.  For their model assumptions regarding a Fermi gas level
density, negligible emission barriers and a linear increase of available exit
channels with increasing excitation energy, they concluded that the ratio should
be constant with a value near 2. They also concluded that the statistical
variance of this ratio would be small enough to enable this ratio to be used as
an identifier of statistical de-excitation on an event by event
basis~\cite{horn1996signatures}.  Experimental observations of constant values
of the ratio have been cited as evidence for strong dominance of statistical
de-excitation of projectile-like fragments~\cite{larochelle1996dependence,
colin2003dynamical, baran2004neck, larochelle1997formation,
fuchs1994heavy,horn1996signatures}.  In Fig.~\ref{fig:fig3} we present, for all
observed PLF exit channels with 10 or more events having A~=~40 and Z~=~20 (not
including n or p as discussed above), a plot of average excitation energy, \ex,
vs exit channel separation energy, -Q. In general these data are similar
to previous results~\cite{larochelle1996dependence, horn1996signatures,
beaulieu1994breakup}. A linear fit to these data leads to a slope parameter of
2.39. This result, well above 2, is close to that extracted in
reference~\cite{larochelle1996dependence}. Based upon comparisons with
statistical model results the authors of
reference~\cite{larochelle1996dependence} concluded that there are important
dynamic effects in mid peripheral and central reactions at~35~MeV/nucleon and
above. A closer investigation of Fig~\ref{fig:fig3} indicates that some prominent channels, particularly at lower separation energies, 
have ratios well above the average values. This observed deviation suggests that these reactions warrant additional exploration. 
We return to these results in the section~\ref{sec:alpha_channels2}.
\section{\label{sec:alpha_channels}Alpha-Conjugate Exit Channels} 
\begin{figure} 
\includegraphics[width=0.9\linewidth]{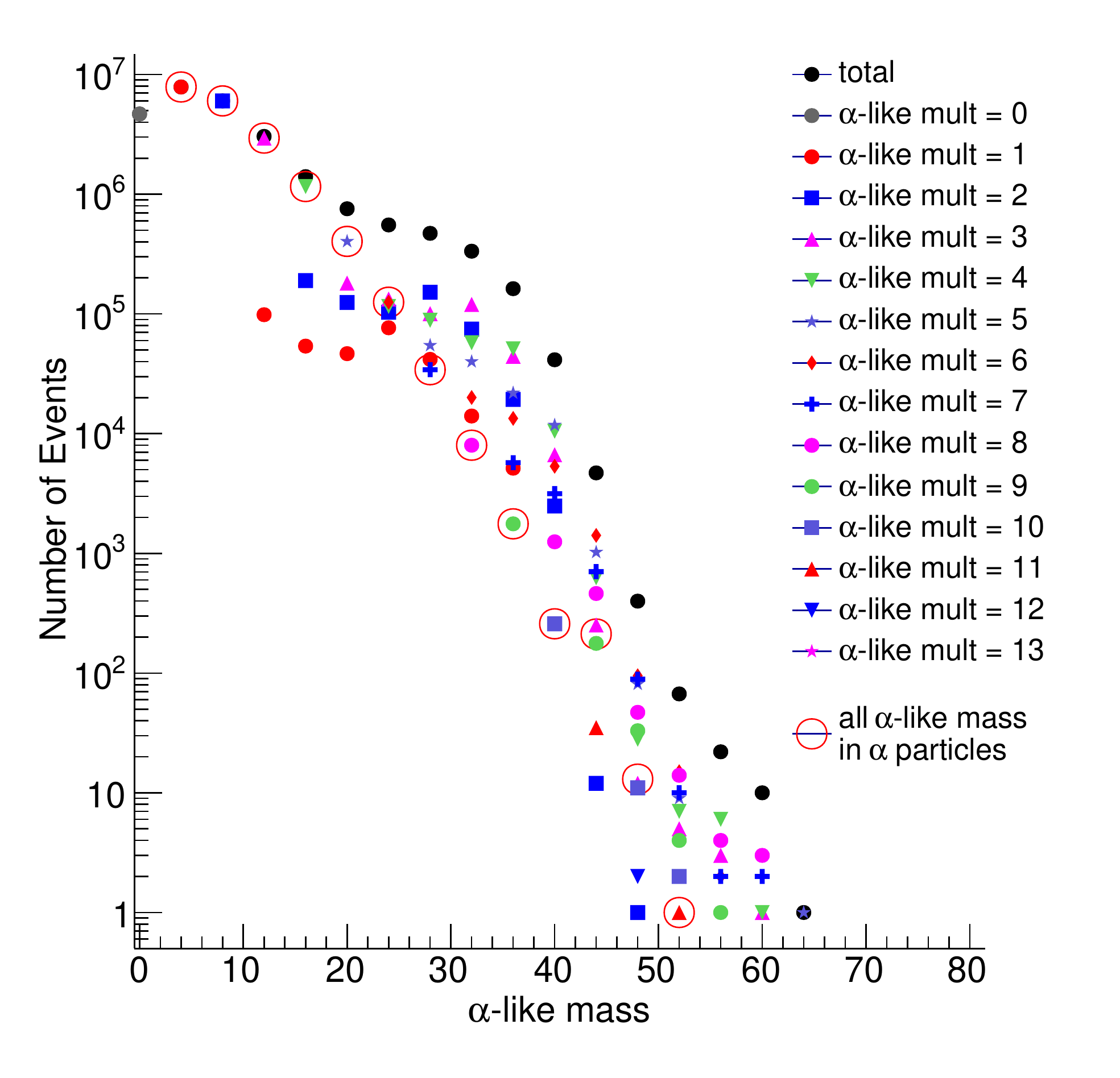}
\caption{\label{fig:fig4}}(Colour online.) Detected number of events yielding
\als particles or
\al-conjugate nuclei in the collision of \cas with \cas at 35A MeV, plotted
against total detected mass of \al-conjugate nuclei. Small filled  circles
represent total yields. Open circles represent yields for events in which only
\als particles contribute to the \alm.
\end{figure}

The main
purpose of the present study was to explore exit channels composed of \als
particles or \al-conjugate nuclei. To focus on such channels the event by event
data were sorted as a function of the total detected ``\al-like mass'', \alm,
i.e., the sum of the masses of the detected products that are either \als
particles or \al-conjugate nuclei. Fig.~\ref{fig:fig4} depicts the resultant event yields.
For a given total \al-like mass, several different decay channels are often
possible. Events for which all of the detected \als conjugate mass is in \als
particles are indicated by the large open circles in Fig.~\ref{fig:fig4}. A total
\al-like mass as large as 85\procs of the entrance channel mass is seen, but with
very low statistics.  The shoulder in the \alm$\thicksim$40 region and rapid decrease beyond
that reflects the detector selectivity for projectile-like fragments from
mid-peripheral events. 
\section{\label{sec:alpha_channels2}Alpha-Conjugate \alm~=~40  Exit Channels}
\begin{figure} 
\includegraphics[width=0.9\linewidth]{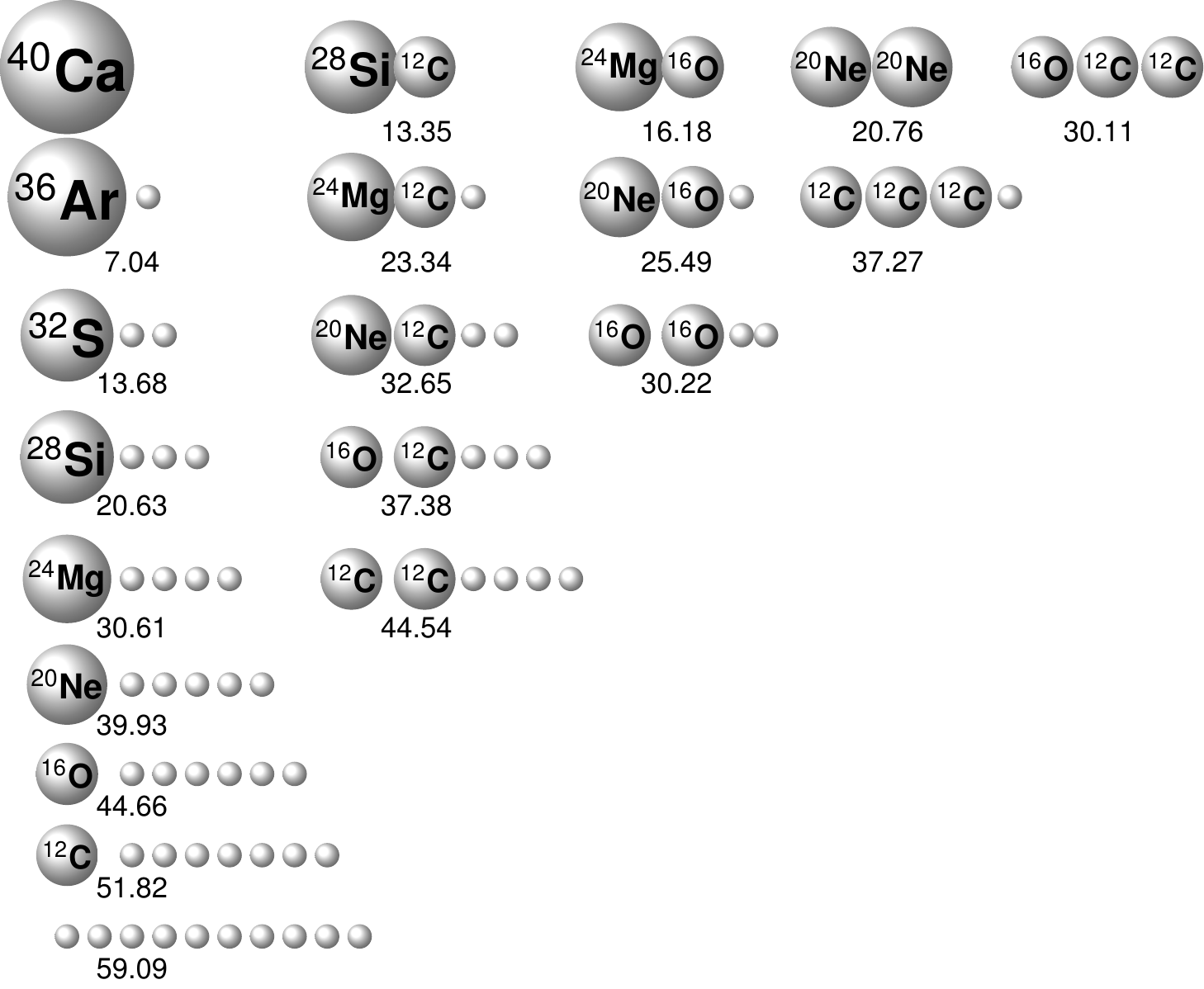}
\caption{\label{fig:fig5}}Ikeda-like diagram for the possible  \al-conjugate
components of \ca. Separation energy (-Q) in MeV for each decay channel is shown. 
\end{figure}

\begin{figure} 
\includegraphics[width=\linewidth]{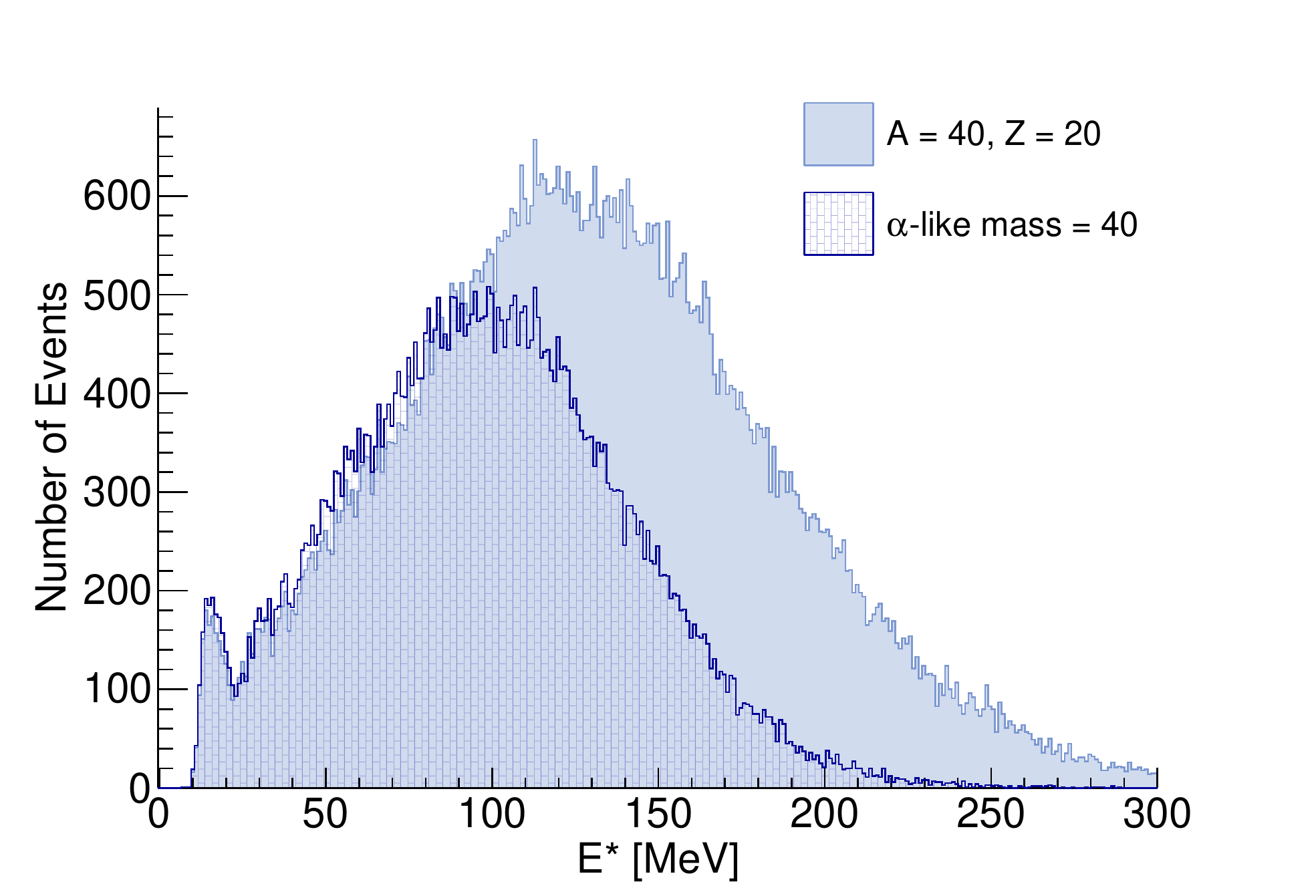}
\caption{\label{fig:fig6}}(Colour online.) Excitation energy distributions for A~=~40, Z~=~20 derived
as indicated in the text, The blue  area represents the data for
all such events. The hatched area represents the data for \alm~=~40 events.
\end{figure}

\begin{figure} 
\includegraphics[width=\linewidth]{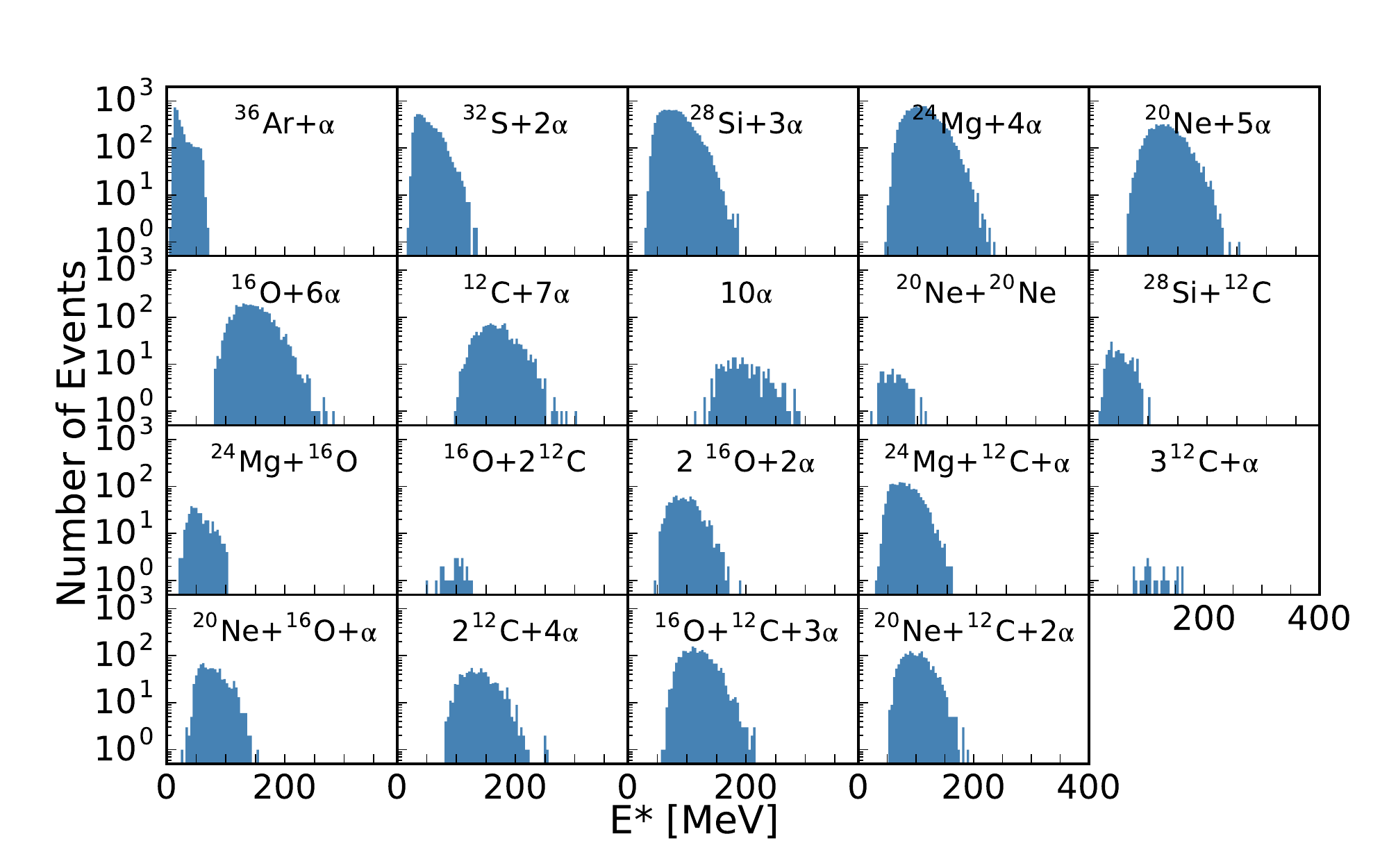}
\caption{\label{fig:fig7}}Excitation functions for the \alm~=~40 events discussed
in the text. 
\end{figure}

\begin{figure} 
\includegraphics[width=\linewidth]{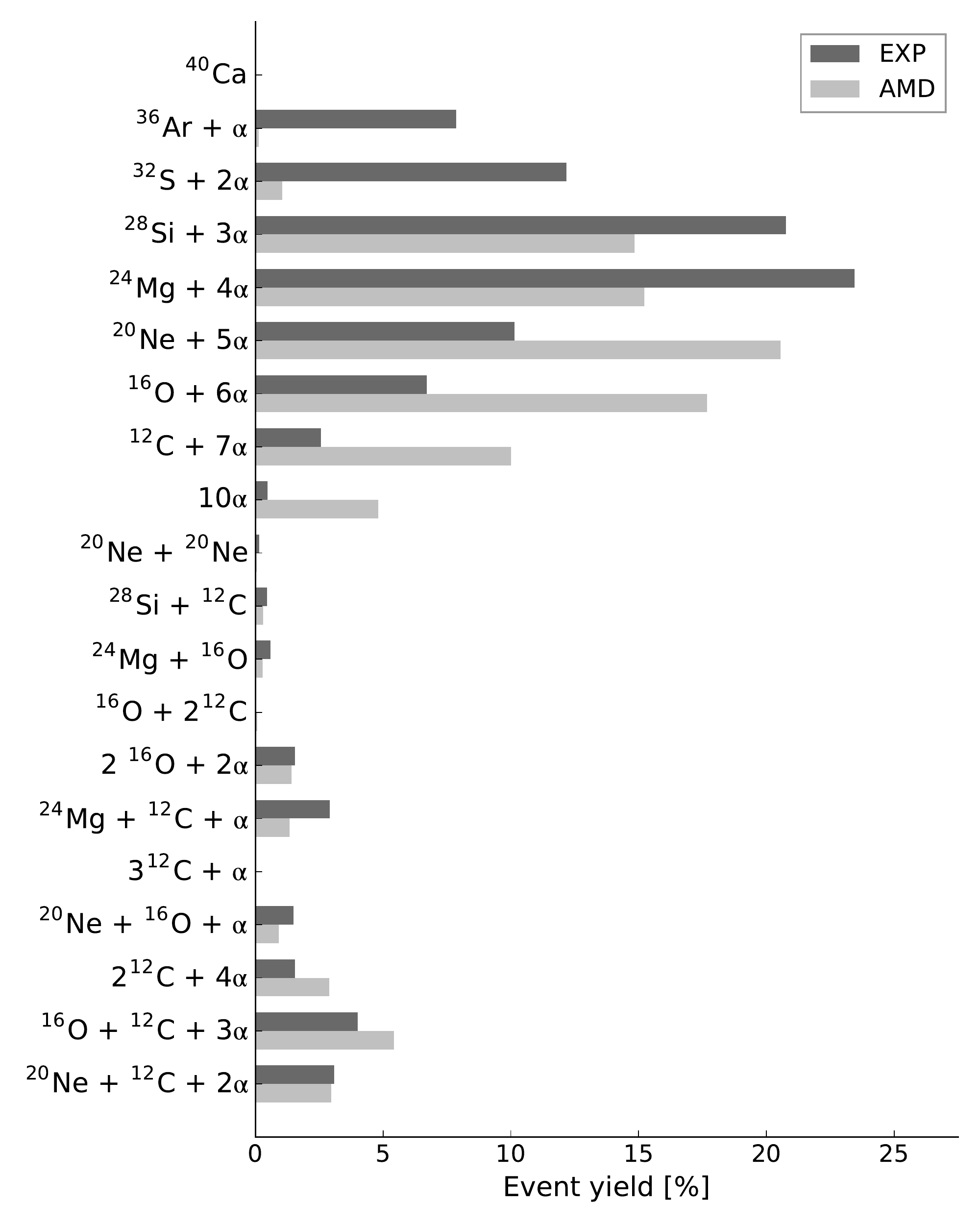}
\caption{\label{fig:fig8}}Percentages of \alm~=~40 events appearing in the
possible exit channels.  The   experimental results are represented by solid
dark grey bars. The filtered AMD-GEMINI results are represented by solid silver bars.
\end{figure}

For the analyses which follow we have chosen to focus on those events for which
\alm~=~40 and compare the properties of the 19 possible exit channels for the
disassembly of the \cas nucleus into \als particles or \al-conjugate nuclei.
The 19 possible combinations of \als-conjugate nuclei which satisfy this total
\al-conjugate mass~=~40 criterion are schematically indicated in
Fig.~\ref{fig:fig5}.
This depiction is similar to that of the Ikeda diagram which is commonly invoked
in discussions of the cluster structure of light
nuclei~\cite{horiuchi1968molecule}.  
The events selected typically have a few Z~=~1 particles (and neutrons)  and, in
rare cases, a heavier non-\al-conjugate fragment, associated with them. To
further refine our event selection we exclude the fraction of the
\alm~=~40 events 
(11\proc) with non-\al-conjugate fragments from the analysis. In
our selection we have allowed Z~=~1 particles and neutrons but we have
re-determined the excitation energies by excluding the Z~=~1 particles and
neutrons as they are primarily pre-equilibrium particles, representing energy
dissipation but not energy deposition into the PLF~\cite{larochelle1999probing}. 
This leads to slightly
smaller excitation energies and introduces a small uncertainty.  As invariant
velocity plots for the \als particles indicate that a small fraction of the
\als particles may result from pre-equilibrium emission or from the target like
source, \als particles with PLF source frame energies greater than 40 MeV were
also excluded to remove those contributions.
The excitation energy distributions derived for all A~=~40, Z~=~20 PLF events
defined in this manner are presented in Fig.~\ref{fig:fig6}.  The \alm~=~40 
events account for
61\procs of the A~=~40, Z~=~20 PLF events detected. Detected events with \al-conjugate
mass~=~40 account for to 0.23\procs of the total experimental events collected.
Filtered AMD calculations predict about half that amount, 0.11\proc.
In Fig.~\ref{fig:fig7} the excitation functions for the different \alm~=~40 exit channels
detected in this reaction are presented. 
The distribution of yields in the different exit channels are presented in
Fig.~\ref{fig:fig8} as percentages of the total \alm~=~40 yields. Both the experimental
results and those from the filtered AMD-GEMINI calculation are presented. 
\begin{figure} 
\includegraphics[width=0.45\textwidth]{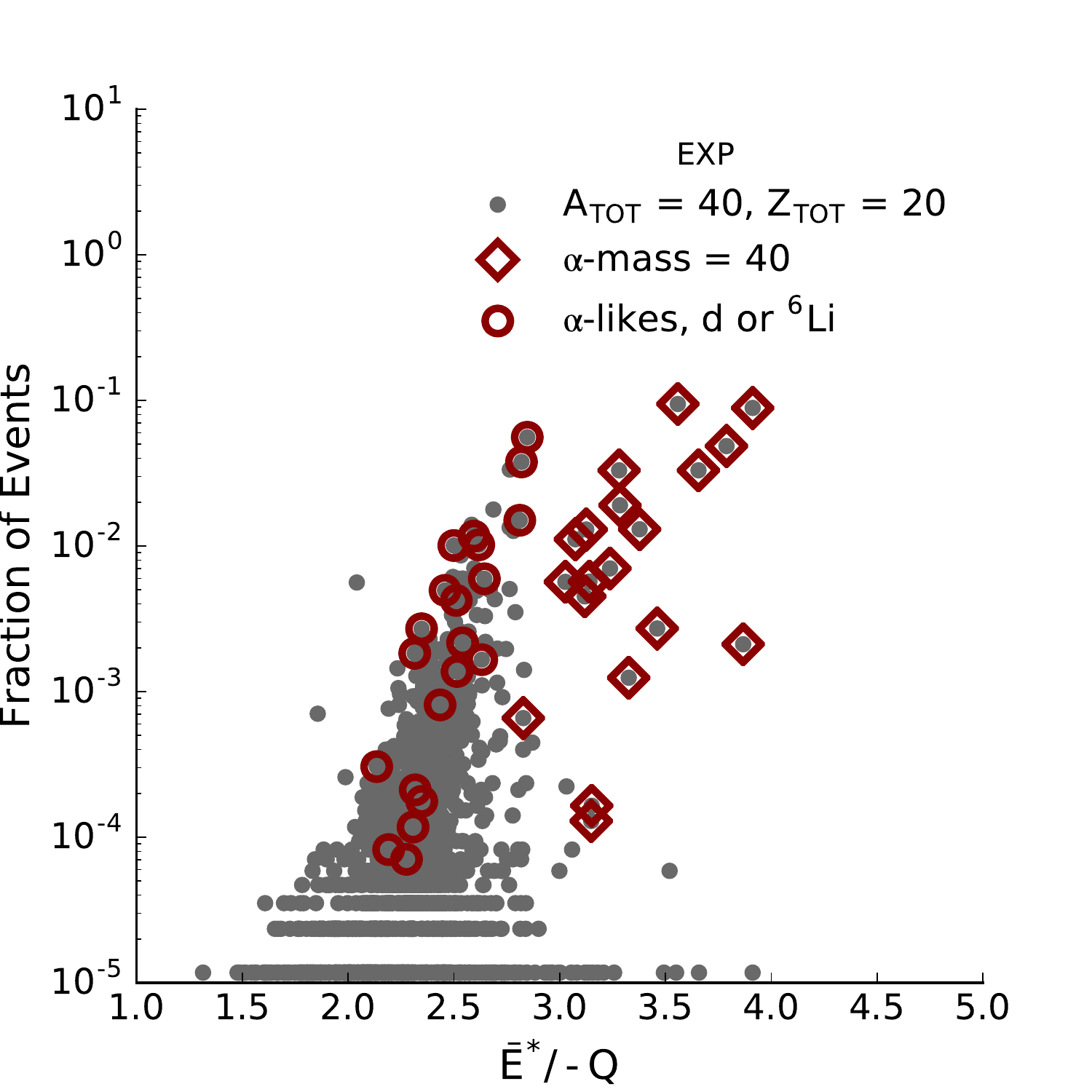}
\includegraphics[width=0.45\textwidth]{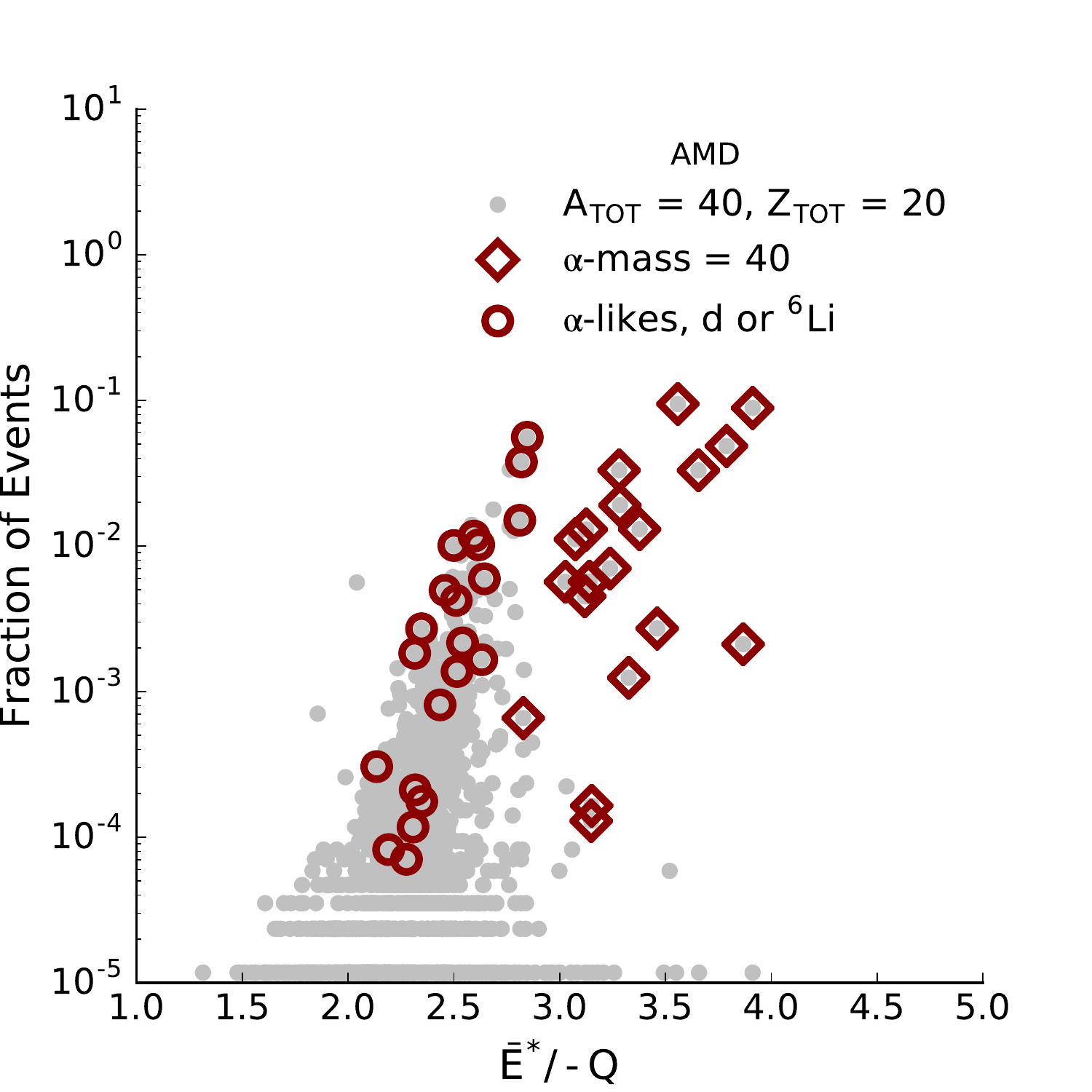}
\caption{\label{fig:fig9}}(Colour online.) The fraction of exit channel events as a function of
the ratio of the average excitation energy to the separation energy. Top -
Experimental data, Bottom - AMD Calculation. Each solid circle represents an exit
channel, The \alm~=~40 channels are identified using large open diamonds. Events
identified by open circles may be \alm~=~40 channels which have undergone
secondary decays with d and \lis emissions (See text).
\end{figure}
They both suggest that the most
probable decay modes are those with one heavy \al-like mass fragment and several
\als particles in the exit channel. While the two distributions are similar, there
are some significant differences between the experimental and calculated
results. In Fig.~\ref{fig:fig9} we plot, for each identified exit channel of the decay of the
selected A~=~40, Z~=~20 nuclei, the fractional yield  vs  the ratio of average
excitation energy to exit channel separation energy (see Fig.~\ref{fig:fig3}). Results are
presented for both the experimental data (top) and AMD simulation (bottom). Each
exit channel is represented by a solid circle. 
We have further identified the \alm~=~40 exit channels using open diamonds. We
see that, in both frames of Fig.~\ref{fig:fig9}, these channels are those with the largest
values of \ex/-Q from the systematics. Their ratios
are well above the values for the other channels with similar separation
energies and in general their yields are quite high. 
An exploration of other high yield channels reveals that
these are generally channels in which the deviations from \alm~=~40 reflect the
existence of deuterons or \lis nuclei in the exit channel. These are exit
channels such as (\fos,2\al,d),(\glin,3\al,d), (\na,4\al,d) and (\na,\li,3\al) for
example. We identify such channels with additional open circles around the
solid circles. These channels might well be those in which an initial breakup
into \als particles and/or \al-conjugate fragments is followed by a secondary
emission or break-up. If so, the fraction of initial \al-conjugate break-ups
of A~=~40, Z~=~20 nuclei is much larger than the 61\procs observed in
Fig.~\ref{fig:fig6}. The
excitation energy evolution of Figs.~\ref{fig:fig1} and~\ref{fig:fig2} suggest 
that it is the dynamic
evolution which favors the extension of these excitation functions to higher
energies and shifts the ratios higher. The degree to which this large fraction
of \al-conjugate de-excitations reflects the initial \al-conjugate nature
of \cas or the dynamic evolution of the excitation and density  warrants further
investigation.  

\section{\label{sec:dyn}Collision Dynamics for \alm~=~40}
\begin{figure}
\includegraphics[width=\linewidth]{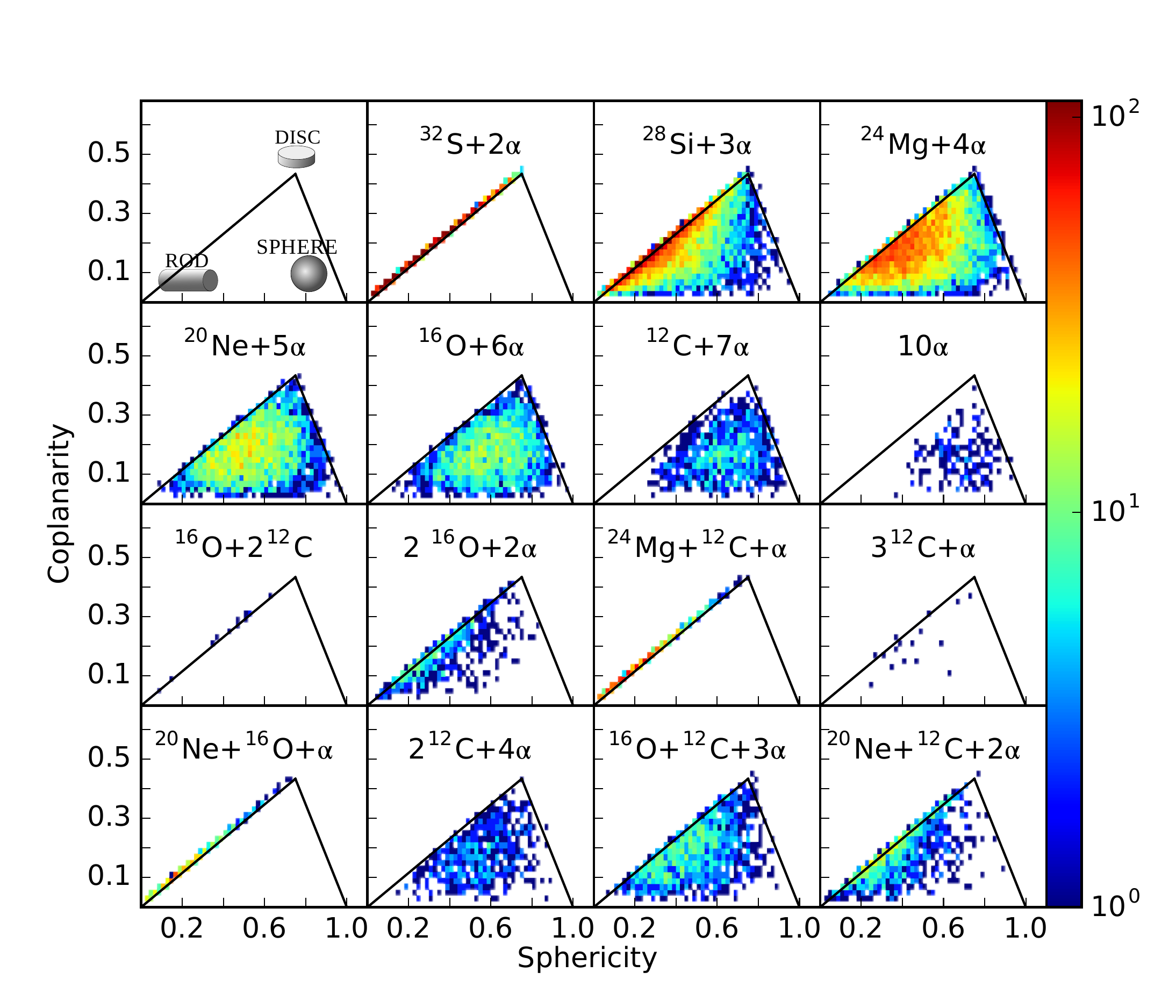}
\caption{\label{fig:fig10}}(Colour online.) Sphericity Co-planarity plots for the \alm~=~40 exit
channels.  Two-body exit channels are excluded. See text. 
\end{figure}

\begin{figure} 
\includegraphics[width=0.5\textwidth]{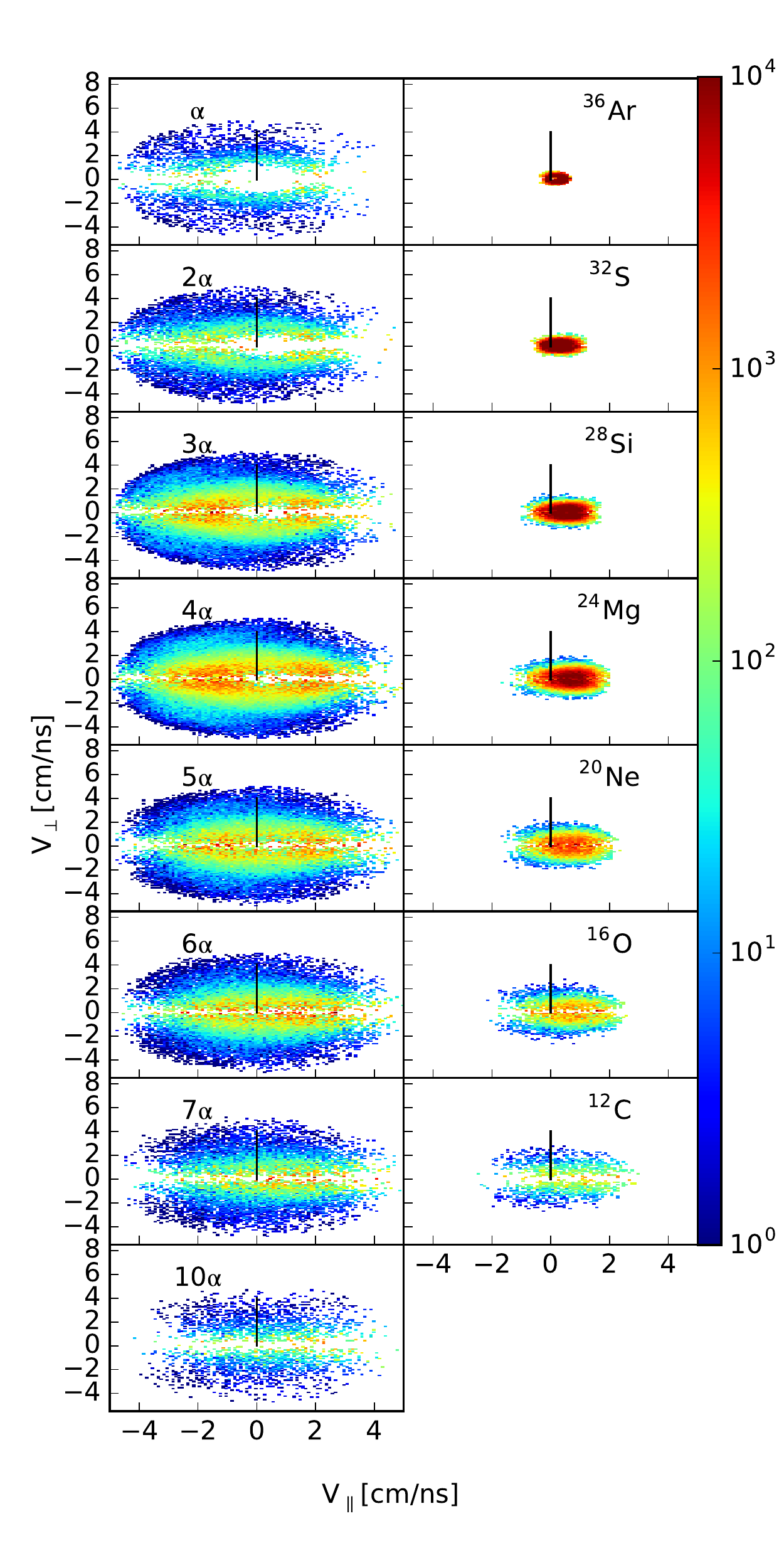}
\caption{\label{fig:fig11}}(Colour online.) Source frame invariant velocity plots for the \al-like
exit channels containing \als particles. Vertical lines at 0 are to aid the
eye in comparisons of these distributions. 
\end{figure}

\begin{figure} 
\includegraphics[width=0.5\textwidth]{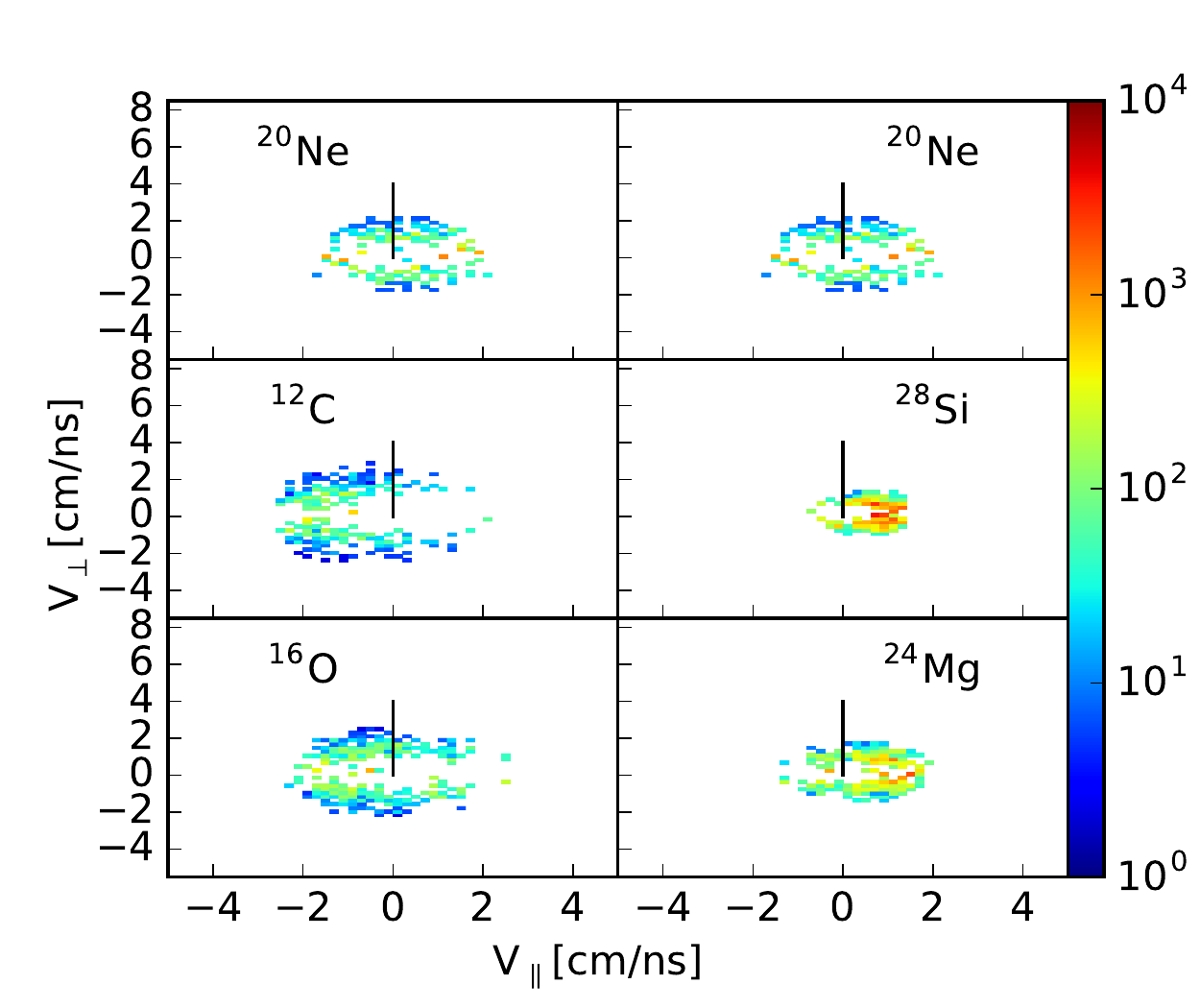}
\caption{\label{fig:fig12}}(Colour online.) Source frame invariant velocity plots for the two
fragment \al-like exit channels.  Vertical lines at 0 are to aid the eye in
comparisons of these distributions.
\end{figure}

To more explicitly probe the dynamics of the \alm~=~40 events we have constructed
momentum space representations of the correlations among exit channel products
using sphericity and co-planarity to characterize  the event
shapes~\cite{fai1983statistical, bondorf1990finiteness}. Sphericity, S, and  Coplanarity, 
C, are defined as:
\begin{eqnarray} 
S = \frac{3}{2}\frac{\lambda_1 + \lambda_2}{\lambda_1 + \lambda_2 + \lambda_3},
\label{eq:sph} 
\end{eqnarray} 

\begin{eqnarray} 
C = \frac{\sqrt{3}}{2}\frac{\lambda_2 - \lambda_1}{\lambda_1 + \lambda_2 +
\lambda_3}.
\label{eq:copl} 
\end{eqnarray} 
Where the $\lambda$ are the eigenvalues of the flow tensor in the c.m. of the source
system and are ordered so that $\lambda_1 < \lambda_2 < \lambda_3$. 
A combined plot of S and C reveals the dominant shape in momentum space.\\
\indent The upper left panel of Fig.~\ref{fig:fig10} provides a schematic representation of the
interpretation of momentum space distributions using these coordinates. Events
at 0.0, 0.0 are rod-like. Those at 0.75, 0.43 are disk like. Events along the line
between these points are co-planar. The events at 1.0, 0.0 are spheres. Oblate and
prolate shapes will appear in the regions between these extremes. It is
important to note that the shapes in the sphericity-coplanarity plots do not
reflect the actual geometric shape of the decaying nuclei, but they represent
the shape of the momentum flow during the decay.
In the rest of Fig.~\ref{fig:fig10} we present the experimental sphericity - coplanarity plots for the \alm~=~40 exit channels. 
We do not include the channels with only two \al-conjugate fragments which would necessarily appear at 0.0 in the sphericity-coplanarity 
plane. 
Most of exit channel event distributions fall closer to the co-planar region of
the rod to disk axis than that of the sphere and only the larger multiplicity
events approach the latter. In some previous work similar observations have been
attributed to multiplicity effects~\cite{bondorf1990finiteness}. While it is obvious 
that fluctuations will be important and that two and three fragment events will 
necessarily be
co-planar in this representation, in general, the distribution will reflect the
initial momentum distribution resulting from the collision  as well as  the mode
and sequence of subsequent de-excitations and momentum conservation in that
sequence rather than the multiplicity, per se. The generally prolate nature of
the sphericity co-planarity plots of Fig.~\ref{fig:fig10} suggest that the exit channels
with large numbers of \als particles result from processes in which an initial
breakup into larger excited  fragments is followed by \als particle
de-excitation. Under very specific circumstances of simultaneous fragmentation,
the observed momentum space shape should be more directly related to the initial
geometric configuration of the de-exciting system~\cite{najman2015freeze}.\\
To understand these \alm~=~40 events in more detail, we have constructed invariant
velocity distributions for the single fragment (x\al) exit channels,
Fig.~\ref{fig:fig11}, and the two fragment exit channels, Fig.~\ref{fig:fig12}.
The products of the different
decay channels are transformed into the rest frame of the reconstructed
\al-like mass 40 nucleus. The decay channels are indicated in the various
panels. The vertical lines indicate the rest frame parallel velocity of the
reconstructed emitting source. In these figures the right-hand panels show the
invariant velocity distributions for the heaviest fragment in the event and the
left-hand panels show the invariant velocity distribution for the \als
particles or other remnants of the de-excitation. In Fig.~\ref{fig:fig11} we note that
the velocity spectra of the heaviest fragment is peaked at a parallel velocity
above the reconstructed source  velocity while the \als particle velocities
are centered  at lower parallel velocities than the reconstructed source
velocity. We also note that the \als particle velocity distributions become
more symmetric about the source velocity as the multiplicity of \als particles
increases. In Fig.~\ref{fig:fig12} we show the decay channels of \al-like mass 40 nuclei 
which consist of pairs of heavier \al-like mass fragments. Except for the symmetric two \neons channel, 
we observe a similar behavior - the heavier fragment velocities are centered at velocities larger 
than the velocity of the decaying nucleus while the velocity distributions for the lighter fragments 
peak at parallel velocities smaller than the parallel velocity of the decaying nucleus.   

\begin{figure*} 
\includegraphics[width=0.85\textwidth]{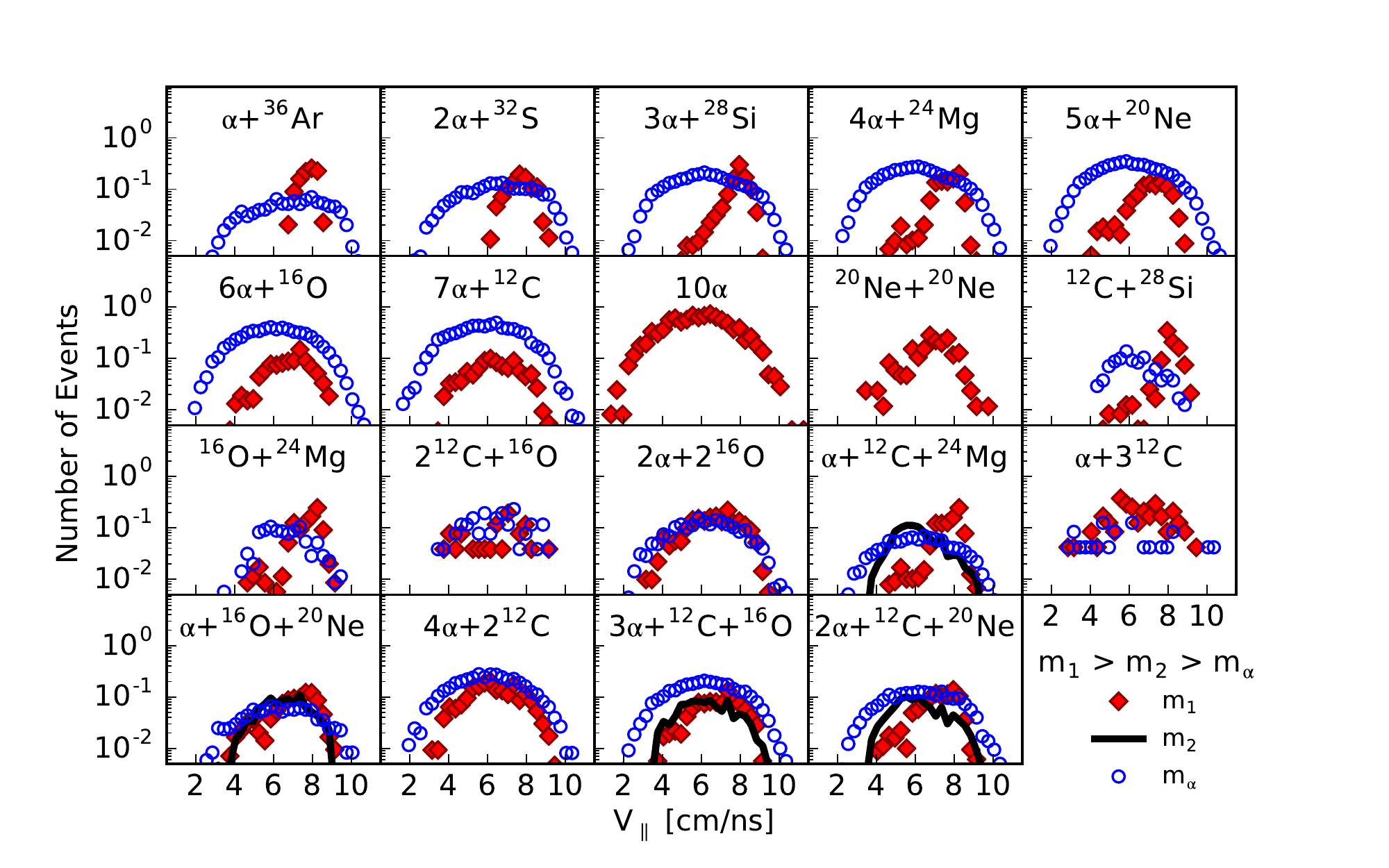}
\caption{\label{fig:fig13}}(Colour online.) Parallel velocity distributions for the
\al-conjugate exit channels. In each panel distributions are color coded for
different products. Solid red diamonds: heaviest fragment in the event, black lines: second heaviest
fragment, open blue circles: \als particles.
\end{figure*}

To emphasize the generality of this observation for the \alm~=~40 exit channels,
we show distributions of observed mass vs parallel velocity for all the
different decay channels in  Fig.~\ref{fig:fig13}. We note again in this figure that the
heaviest fragment in the different decay channels always tends to be observed at
velocities larger than that of the neck region and that light particles tend to
be observed as originating from the  velocity region between the  source
velocity and the COM velocity, i.e., from a neck region. To verify that the
effect is real and not the result of some biasing by the experimental acceptance
of NIMROD, we have done statistical model calculations using the statistical
de-excitation code GEMINI. When these events were filtered through our
experimental acceptance the resultant parallel velocity distributions remained
symmetric about the source velocity. 
In the present case these necks exhibit important \als clustering effects. The
manifestation of this neck can be either a single \al-conjugate fragment or
one or more \als particles either independently formed or derived from the
de-excitation of an excited \al-conjugate precursor. The observed emission patterns, in which the lighter fragments trail the heavier fragments, 
are strongly reminiscent of the ''hierarchy'' effect reported for other systems in a similar energy range~\cite{colin2003dynamical, baran2004neck}. 
It reflects a dynamics in which mass and velocity are correlated such that, for fragments emitted forward in the center of mass, the heaviest fragments 
are emitted at forward angles and are on average the fastest ones, the second heaviest fragment is the second fastest one, and so on.  
Such behavior is inconsistent with production of a fully equilibrated compound nucleus. Rather it signals a binary nature of the reaction with neck 
formation between the quasi-projectile and the quasi-target~\cite{baran2004neck, larochelle1997formation}. The breakup of this neck is fast enough 
that memory of the neck geometry is retained. Of course these emissions from the neck region are subject to possible modification by proximity 
effects~\cite{charity1995prompt, hudan2004interplay, hudan2006short, mcintosh2007tidal, montoya1994fragmentation}.
The results in Figs.~\ref{fig:fig11}~-~\ref{fig:fig13} suggest that the \als particles
in x\als events
observed in the left panels of Fig.~\ref{fig:fig11} could originate from the same
process as the fragments seen in the left-hand side of Fig.~\ref{fig:fig12}. As previously
noted, the \cas+\carbs reaction at 25~MeV/nucleon populates excited states of
\carbs
nuclei which decay by 3\als emission, primarily in a sequential
manner~\cite{raduta2011evidence}.
It is reasonable to expect that similar excited \als de-exciting states are
produced in the present reaction. Indeed, we have already noted that the
sphericity co-planarity plots of Fig.~\ref{fig:fig10} suggest that the exit channels with
large numbers of \als particles result from processes in which an initial
breakup into larger excited fragments is followed by \als particle
de-excitation. Further evidence for such precursors is found in our data in the
large numbers of $^8$Be nuclei emitted. The granularity of the detector in our
experiment is such that most of these are observed as two \als particles
simultaneously striking a single detector and identified by their combined
$\Delta$E, E
signal. The granularity of our detector is not well suited to measuring the $^8$Be
correlation function so we do not pursue this question further.

\section{\label{sec:concl}Summary and Conclusions}
Reactions of 35 MeV/nucleon \cas with \cas have been investigated with an
emphasis on peripheral and mid-peripheral collisions leading to excited
projectile-like fragments.  A global analysis of the de-excitation channels of
A~=~40 PLF fragments agrees with previous studies that total equilibration of all
degrees of freedom is not achieved in the mid-peripheral collisions. A hierarchy
effect is observed in the collision dynamics. The selection of the subset of
A~=~40 projectile-like fragment exit channels characterized by a total
\al-conjugate mass (\als particles plus \al-conjugate fragments) equal to
40 indicates that these projectile-like exit channels generally have important
dynamic contributions. Most of the \als particles observed in such events trail
larger \al-conjugate leading fragments and originate from \al-conjugate neck
structures formed during the collisions. The manifestation of this neck can be a
single \al-conjugate fragment or one or more \als particles either
independently formed or derived from the de-excitation of an excited \al-
conjugate precursor. This mechanism significantly increases the difficulty of isolating 
clean projectile decay samples~\cite{montoya1994fragmentation}\\
Transport model calculations typically indicate that the neck structures formed
in mid-peripheral collisions have densities lower than normal
density~\cite{baran2004neck}.
Lowering of the density is expected to favor \als clustering. Using a
constrained HFB approach, Girod and Schuck have explored the nuclear equation of
state for self-conjugate N=Z nuclei and concluded that those nuclei will
cluster into a metastable phase of \als particles (or in some cases \al-conjugate 
light clusters) at excitations above ~ 3 MeV/nucleon and densities
below ~0.33 normal density~\cite{girod2013alpha}. We believe that the reaction dynamics observed
in this paper can provide a natural entry point to study the disassembly of
\als clustered systems with potentially exotic geometries and properties.\\
\begin{acknowledgments}  
This work was supported by the United States Department of Energy under Grant
$\#$DE-FG03- 93ER40773 and by The Robert A. Welch Foundation under Grant $\#$A0330. We
appreciate useful conversations with S. Shlomo. S. Umar and A. Ono. We also
greatly appreciate the continued excellent work of the staff of the TAMU
Cyclotron Institute. 
\end{acknowledgments}
\bibliography{HierarchyKSchmidt}
\end{document}